\documentclass[twocolumn,showpacs,preprintnumbers,amsmath,amssymb]{revtex4}

\usepackage{color}
\usepackage{graphicx}
\usepackage{dcolumn}
\usepackage{bm}

\begin{document}


\title{Absorption-free superluminal light propagation in
a V-type system}

\author{Kh. Saaidi}
 \email{ksaaidi@uok.ac.ir}
\author{B. Ruzbahani}%
\author{S. W. Rabiei}%
 \email{w.rabiei@uok.ac.ir}
\affiliation{%
Physics Department, University of Kurdistan\\
Sanandaj, Iran}%

\author{M. Mahmoudi}
\email{mahmoudi@iasbs.ac.ir}
\affiliation{
Physics Department, Zanjan University, P. O. Box 45195-313, Zanjan, Iran
}%

\date{\today}

\begin{abstract}
Dispersion and absorption properties of a weak probe field in a three-level V-type atomic system is studied. By application of indirect incoherent pump fields the effect of populating upper levels on optical properties of the atomic medium in presence of a strong coherent pump field is investigated. It is shown that the slope of dispersion switches from positive to negative just by changing the intensity of the coherent or indirect incoherent pump fields. It is demonstrated that the absorption-free superluminal and subluminal light propagation appear in this system.
\end{abstract}

\maketitle

\section{Introduction}\label{sec:level1}

In the past few decades, there has been tremendous interests in the study of subluminal and
superluminal light propagation \cite{1,2,3,4,5}. The group
velocity of light pulse can be reduced in Bose-Einstein condensate of sodium atom gas \cite{1}, and in hot gases \cite{3}, and even halted in vapor of Rb atoms \cite{4}. Also, it can exceed the vacuum light speed, c, and can even become negative \cite{5}. These experiments are based on the fact that electromagnetically induced transparency (EIT) and electro magnetically induced absorption (EIA) \cite{6} lead to a dispersion profile with a sharp positive or
negative derivative \cite{7,8}. Anomalous dispersion was first studied in mechanical oscillators \cite{9} and was later applied by Sommerfeld and Brillouin \cite{10} to light propagation in absorptive opaque materials. They showed theoretically that inside an absorption line, the dispersion can be anomalous, resulting in a group velocity faster than c, the vacuum speed of light. Such an anomalous velocity appears due to the wave nature of light \cite{11,12}. Talukder et al. have shown femtosecond laser pulse propagation has switched from superluminal to subluminal velocities in an absorbing dye by changing the dye concentration \cite{13}. Shimizu et al. were able to control the light pulse speed with only a few cold atoms in a high-finesse microcavity by detuning the laser frequency from a cavity resonance frequency-locked to
the atomic transition \cite{14}. In a series of papers \cite{15,16,17,18,19,20,21,22}, Chiao and coworkers showed theoretically that anomalous dispersion can occur inside a transparent material. It was predicted that by using a gain doublet \cite{18}, it is possible to obtain a transparent anomalous dispersion region where the group velocity of light pulse exceeds c with almost no pulse distortion. Wang et al. used gain-assisted linear anomalous dispersion in atomic cesium gas, and the group velocity of a laser pulse in their experiment exceeded c and could even become negative, while the shape of the pulse was preserved \cite{5}. The incoherent pumping fields can also play an important role in the controlling of the group velocity of light in dispersive media \cite{23,24}. The double-$\Lambda$ setup is another scheme which provides a very rich spectrum of phenomena based on atomic coherence \cite{25}. Recently, the dispersion and the absorption properties of a weak probe field in simple multi-level quantum systems have been considered \cite{26,27,28}. They showed that, by using a coherent and an indirect incoherent pump field the group velocity of light  can be controlled. There have been only few experimental and theoretical studies in which both free absorption superluminal and subluminal light propagation in a single system have been realized.
In this article, the dispersion and the absorption properties of a weak probe field in a three-level V-type atomic system is investigated. By application of indirect incoherent pump fields the effect of populating upper levels on optical properties of the atomic medium in presence of a strong coherent pump field is investigated. It is shown that the slope of dispersion changes from positive to negative just by adjusting
the intensity of the coherent or indirect incoherent pump fields. It is demonstrated that the absorption free superluminal and subluminal light propagation appear in this system.

\section{The model and discussions}\label{sec:level2}
\begin{figure}[h]
\centering
  \includegraphics[width=6.4cm]{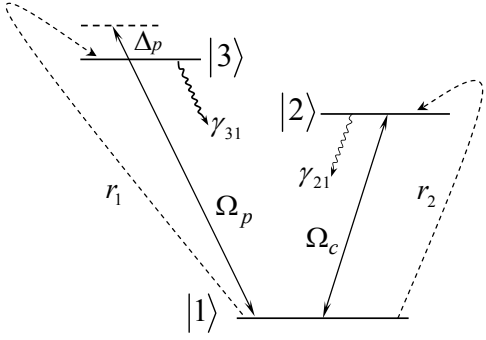}
  \caption{The proposed level scheme. A three-level V-type atomic system driven by a strong coherent coupling and a weak tunable probe field. The indirect incoherent pumps are denoted by the dashed curve lines.}
 \label{fig:F1}
\end{figure}
The model consists of a closed three-level V-type atomic system with ground level $\left| 1 \right\rangle$ and two
upper levels  $\left| 2 \right\rangle$ and $\left| 3 \right\rangle$ as shown in Fig.\ref{fig:F1}. A strong coherent coupling
field with Rabi frequency $\Omega _c  = \vec E_c .\vec \wp _{12} /\hbar$ drives transition $\left| 1 \right\rangle  \leftrightarrow \left| 2 \right\rangle$, while a weak tunable probe field with Rabi frequency $\Omega _p  = \vec E_p .\vec \wp _{13} /\hbar$ is
applied to transition $\left| 1 \right\rangle  \leftrightarrow \left| 3 \right\rangle$. Here, $\vec \wp _{1j}$$(j=2,3)$  are the atomic dipole moments, and $E_c$ ($E_p$) is the amplitude of the coupling (probe) field. By indirect incoherent pumps with rates  $r_1$ and $r_2$   the population of the ground level $\left| 1 \right\rangle$  can be pumped to the exited levels  $\left| 2 \right\rangle$  and  $\left| 3 \right\rangle$  via some unspecified auxiliary states \cite{29,29b}. The spontaneous decay rates from states   $\left| 2 \right\rangle$  and  $\left| 3 \right\rangle$  to the ground level  are denoted by $\gamma_{21}$ and $\gamma_{31}$, respectively. The density matrix equations of motion under the rotating wave approximation and in a rotating frame are:
\begin{subequations}
\label{eq:1}
\begin{eqnarray}
\dot \rho _{11}  &=& i\Omega _p \rho _{31}  + i\Omega _c \rho _{21}  - i\Omega _p^* \rho _{13}  - i\Omega _c^* \rho _{12}  \nonumber\\
 &&+ \gamma _{31} \rho _{33}+ \gamma _{21} \rho _{22}  - (r_1  + r_2 )\rho _{11}
\label{subeq:1a}
\\
\dot \rho _{22}  &=& i\Omega _c^* \rho _{12}  - i\Omega _c \rho _{21}  - \gamma _{21} \rho _{22}  + r_2 \rho _{11}
\label{subeq:1b}
\\
\dot \rho _{33}  &=& i\Omega _p^* \rho _{13}  - i\Omega _p \rho _{31}  - \gamma _{31} \rho _{33}  + r_1 \rho _{11}
\label{subeq:1c}
\\
\dot \rho _{21}  &=& (i\Delta _c  - \Gamma _{21} )\rho _{21}  + i\Omega _c^* \rho _{11}  - i\Omega _c^* \rho _{22} \nonumber\\
&&- i\Omega _p^* \rho _{23}
\label{subeq:1d}
\\
\dot \rho _{31}  &=& (i\Delta _p  - \Gamma _{31} )\rho _{31}  + i\Omega _p^* \rho _{11}  - i\Omega _p^* \rho _{33} \nonumber\\
&&- i\Omega _c^* \rho _{32}
\label{subeq:1e}
\\
\dot \rho _{32}  &=& \left( {i(\Delta _p  - \Delta _c ) - \Gamma _{32} } \right)\rho _{32}  + i\Omega _p^* \rho _{12}\nonumber\\
&&- i\Omega _c \rho _{31},
\label{subeq:1f}
\end{eqnarray}
\end{subequations}
where $\Gamma_{ij}$    are the coherence decay rates given by
\begin{eqnarray}
\Gamma _{21}  &=& \frac{1}{2}(\gamma _{21}  + r_1  + r_2 ),
\nonumber
\\
\Gamma _{31}  &=& \frac{1}{2}(\gamma _{31}  + r_1  + r_2 ),
\label{eq:2}
\\
\Gamma _{32} &=& \frac{1}{2}(\gamma _{21}  + \gamma _{31} ).
\nonumber
\end{eqnarray}
$\Delta _c  = \omega _c  - \omega _{21}$ and $\Delta _p  = \omega _p  - \omega _{31}$ are the detunings of the coupling field and the probe field, respectively. Note that the interference due to the different spontaneous emission channels has been ignored. The response of the atomic system to the applied fields is determined by the susceptibility $\chi$, which is defined as \cite{30}
\begin{eqnarray}
\chi  = \frac{{N\wp _{13}^2 }}{{\varepsilon _0 \hbar }}\frac{{\rho _{31} }}{{\Omega _p^* }}
\label{eq:3}
\end{eqnarray}
where $N$ is the atom number density in the medium. The real and imaginary parts of $\chi$ correspond to the dispersion and the absorption, respectively. The so-called group index, $n_g =c/v_g$, is also introduced where $c$ is speed of light in the vacuum and
\begin{eqnarray}
v_g  = \frac{c}{{1 + 2\pi \chi '(\omega _p ) + 2\pi \omega _p \frac{\partial }{{\partial \omega _p }}\chi '(\omega _p )}},
\label{eq:4}
\end{eqnarray}
is the group velocity of the probe field \cite{5,30}.
Here $\chi '$ is the real part of $\chi$.
\begin{figure}[t]
\centering
  \includegraphics[width=8cm]{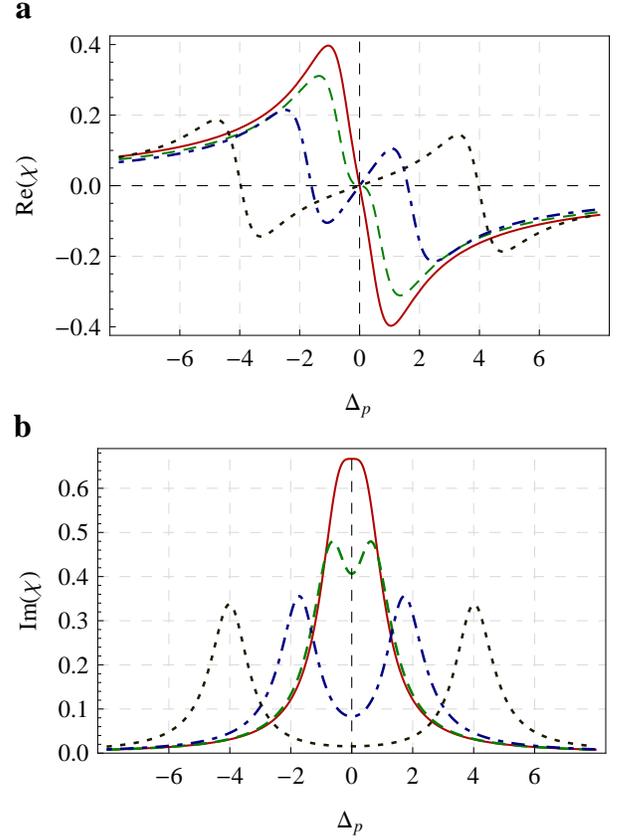}
  \caption{Real (a) and imaginary (b) parts of susceptibility versus probe field detuning for  the parameters $\gamma = 1$, $r_1  = r_2  = 0$, $\Omega_c=0.5$ (solid), $ 0.7$(dashed),  $1.7$ (dot-dashed), and $4.0$ (dotted).}
 \label{fig:F2}
\end{figure}
Eq.(\ref{eq:4}) shows that, when $\chi '$ is negligible the slope of dispersion has the major role in
determination of the group velocity and the group index. In our notation the positive (negative) slope of dispersion corresponds
to the increase (reduction) of the group index and the positive (negative) value of $\chi '' = {\mathop{\rm Im}\nolimits} (\chi)$  shows the attenuation (amplification) of the
probe field. It is apparent that a negative group index means a negative group velocity.
The steady state solutions of Eqs.(\ref{eq:1}) for the weak probe field approximation i.e. $|\Omega _p | \ll \gamma _{ij}$  in the
case of tuned coupling field ( $\Delta _c  = 0$ ) are
\begin{subequations}
\label{eq:5}
\begin{eqnarray}
\rho _{11} &=& \left( {1 + \frac{{r_1 }}{{\gamma _{31} }} + \frac{{2\left| {\Omega _c } \right|^2  + \Gamma _{21} r_2 }}{{\Gamma _{21} \gamma _{21}  + 2\left| {\Omega _c } \right|^2 }}} \right)^{ - 1}
\label{eq:5a}
\\
\rho _{22} &=& \left( {\frac{2\left| {\Omega _c } \right|^2  + \Gamma _{21} r_2 }{\Gamma _{21} \gamma _{21}  + 2\left| {\Omega _c } \right|^2}} \right)\rho _{11}
\label{eq:5b}
\\
\rho _{33} &=& \frac{{r_1 }}{{\gamma _{31} }}\rho _{11}
\label{eq:5c}
\\
\rho _{21}  &=& \frac{{i\Omega _c^* }}{{\Gamma _{21} }}\left( {\rho _{11}  - \rho _{22} } \right)
\label{eq:5d}
\\
\rho _{31}  &=& \frac{{i\Omega _p^* }}{{(\Gamma _{31}  - i\Delta _p ) + \frac{{\left| {\Omega _c } \right|^2 }}{{(\Gamma _{32}  - i\Delta _p )}}}}
\times  \{(\rho _{11}  - \rho _{33} )
\nonumber
\\
&& - \frac{{\left| {\Omega _c } \right|^2 }}{{\Gamma _{21} (\Gamma _{32}  - i\Delta _p )}}(\rho _{11}  - \rho _{22} )\}.
\label{eq:5e}
\end{eqnarray}
\end{subequations}
For simplicity, from this point on, it is assumed that the Rabi frequencies ($\Omega_c$ and $\Omega_p$) are real numbers, the spontaneous
decay rates are the same for
all levels, i.e. $\gamma _{21}  = \gamma _{31} =\gamma$, and other parameters are normalized to the spontaneous decay rate $\gamma$.
In the following, using the derived expression for susceptibility $\chi$, in terms of given system parameters, the response of
the atomic system to the applied coherent and incoherent fields is studied. It has to say that the main interest is in dispersion
and absorption properties of the probe field around zero detuning, $\Delta_p=0$.
\begin{figure}[t]
\centering
  \includegraphics[width=8cm]{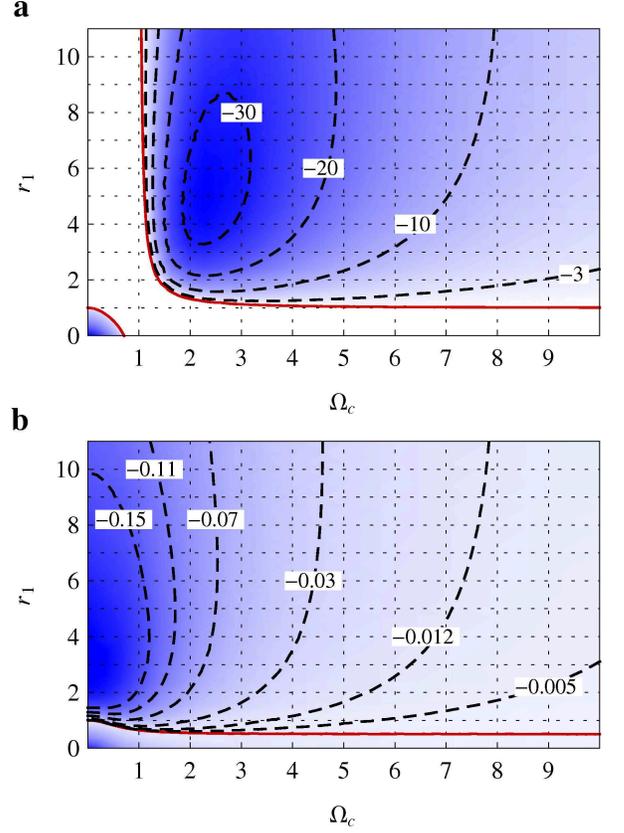}
  \caption{Contour plots for $n_g-1=0$ (red-solid lines), $-30$, $-20$, $-10$, and $-3$ (black-dashed lines) (a) and $\chi''=0$ (red-solid line), $-0.15$, $-0.11$, $-0.07$, $-0.03$, $-0.012$, and $-0.005$ (black-dashed lines) (b) via $\Omega_c$ and $r_1$. The common parameters are $\gamma = 1$, $r_2=0$, $\Delta_c=0$, and $\Delta_p=0$. }
 \label{fig:F3}
\end{figure}
\par
First the familiar effect of the coupling field in absence of the indirect incoherent pumps ($r_1=r_2=0$) is mentioned. In this case
due to application of the coupling field, the absorption line for probe field splits into two absorption lines and consequently
the slope of dispersion changes from negative to positive. In Fig.\ref{fig:F2} the real (a) and imaginary (b) parts of the susceptibility are
plotted versus the probe field detuning. It shows that increasing the value of the Rabi frequency, $\Omega_c$, leads to the appearance
of a transparent region between the absorption lines around zero detuning, $\Delta_p=0$, which is an EIT situation 
proportional
to lossless subluminal light propagation.
For the next step, while the strength of the second indirect incoherent pump is
zero ($r_2=0$ ), the effects of driving transition $\left|1\right\rangle \leftrightarrow \left|2\right\rangle$  by the coupling field
and populating level $\left| 3 \right\rangle$ is investigated.
To get a rough view of this condition, it will be useful to draw some contour plots for $n_g-1$  and $\chi''$, via
the coupling Rabi frequency, $\Omega_c$ , and the first indirect incoherent pump strength,  $r_1$, while $r_2$  and $\Delta_p$  are zero.
Using Eqs.(\ref{eq:5}) and adjusting scale arbitrarily by letting $\frac{{N\wp _{13}^2 }}{{\varepsilon _0 \hbar }} = 1$ and
$2\pi \omega _p  = 10^2 \gamma$, the explicit expressions for  $n_g-1$ and  $\chi''$  in terms of  $\Omega_c$, $r_1$, and $r_2$  at zero probe field detuning ($\Delta_p=0$) are
\begin{eqnarray}
n_g  - 1 &=& \frac{10^2 \gamma}{A^2 B} \left\{{4(r_1  - 1)(r_2  + r_1  + 1)}\right.
\nonumber
\\
&&- 4\left( {r_1 (r_1  - 5) + r_2  + 2r_1 r_2  + r_2^2 } \right)\Omega_c ^2
\nonumber
\\
&&\left.{ - 16(r_1  - 1)\Omega_c ^4 }\right\},
\label{eq:6}
\end{eqnarray}
and
\begin{eqnarray}
\chi '' &=&\frac{{4 \left( {r_2  - 2r_1  + 1} \right)\Omega_c ^2 }}{{AB}}
\nonumber
\\
&&  - \frac{{2 (r_1  - 1)(r_2  + r_1  + 1)}}{{AB}},
\label{eq:7}
\end{eqnarray}
respectively, where
\begin{eqnarray}
A &=& 2\Omega_c ^2  + r_2  + r_1  + 1,
\nonumber
\\
B &=& (r_2  + r_1  + 1)^2  + 4(r_1  + 2)\Omega_c ^2.
\nonumber
\end{eqnarray}
\begin{figure}[t]
\centering
  \includegraphics[width=8cm]{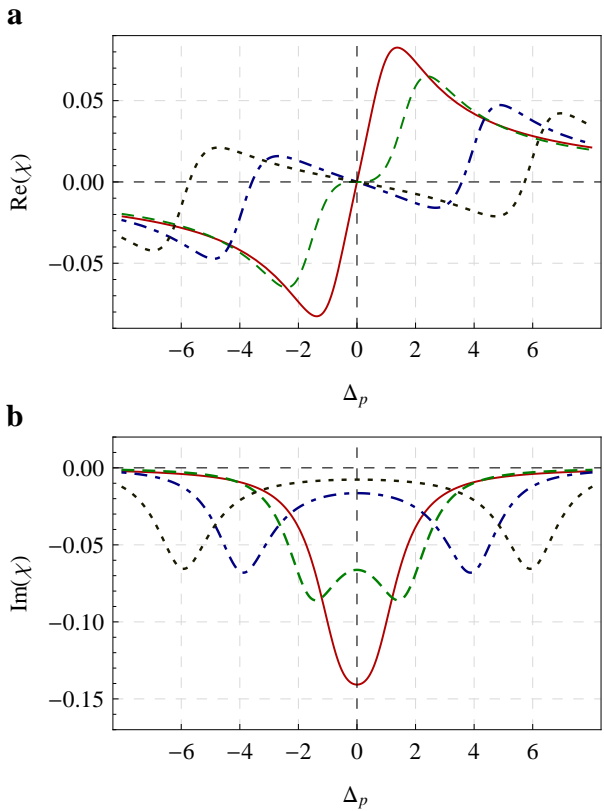}
  \caption{Real (a) and imaginary (b) parts of susceptibility versus probe field detuning for  the parameters $\gamma = 1$, $r_1=1.5$, $r_2=0.0$, $\Omega_c= 0.7$ (solid), $1.7$ (dashed),  $4.0$ (dot-dashed), and $6.0$ (dotted).}
 \label{fig:F4}
\end{figure}
For  $r_2=0$, equations \ref{eq:6} and \ref{eq:7} are used to draw the plot contours of Fig.\ref{fig:F3}a and Fig.\ref{fig:F3}b, respectively.
In Fig.\ref{fig:F3}a, the superluminal regions are colored and the subluminal region is left white. In addition, the
separating contours ($n_g-1=0$) are plotted with red-solid lines. In this figure the contour lines for
($n_g  - 1) = -30, -20, -10$, and $-3$ are plotted in black-dashed style. Fig.\ref{fig:F3}b illustrates the absorption property of the
medium, $\chi ''$. In this figure the absorptive and amplifying regions are separated by the contour line of $\chi '' = 0$ (red-solid line) and
the darkness of the colored regions is a measure of their absorption or amplification magnitude for the probe field. Therefore a
lighter color is used to illustrate the more transparent region. Finally, the contour
lines of $\chi ''= -0.15$, $-0.11$, $-0.07$, $-0.03$, $-0.012$, and $-0.005$, are plotted in black-dashed style.
As Fig.\ref{fig:F3}a illustrates, there are two regions of anomalous dispersion and one region of normal dispersion. As determined by Fig.\ref{fig:F3}b, the first
region of negative dispersion is due to the existence of an absorption line around zero probe field detuning and
occurs where $\Omega_c ,r_1 <1$. The solid lines of Fig.\ref{fig:F2} can be regarded as an example of this region. The other region of
negative dispersion happens where $\Omega _c,r_1> 1$, while according to Fig.\ref{fig:F3}b it accompanies by gain. For the region of normal
dispersion, as Fig.\ref{fig:F3}b demonstrates, the absorption property can take either a small positive or a large negative value, proportional
to negligible attenuation or strong amplification of the probe field, respectively.
\begin{figure}[t]
\centering
\includegraphics[width=8cm]{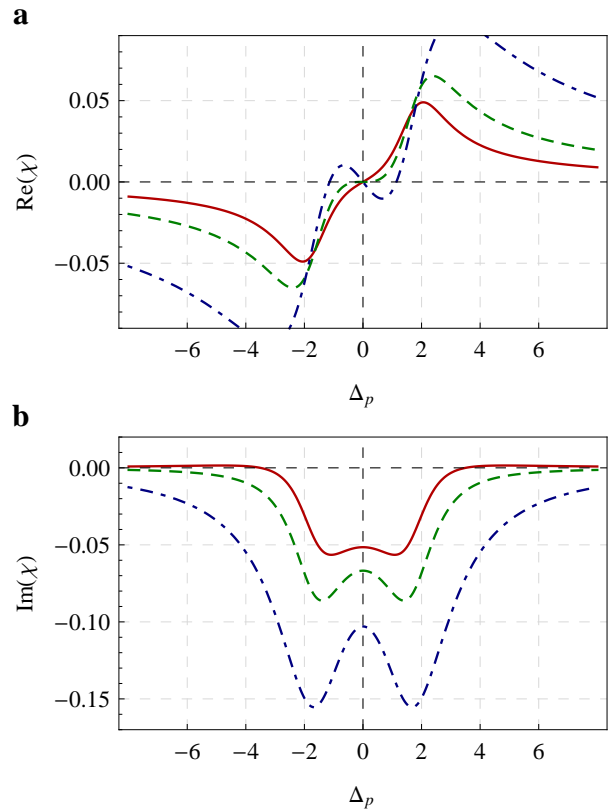}
\caption{Real (a) and imaginary (b) parts of susceptibility versus probe field detuning for the parameters $\gamma = 1$, $ \Omega _c  = 1.69$, $ r_2  = 0$, $r_1=1.2$ (solid), $1.5$ (dashed), and $3.0$ (dot-dashed).}
\label{fig:F5}
\end{figure}
\begin{figure}[t]
\centering
\includegraphics[width=8cm]{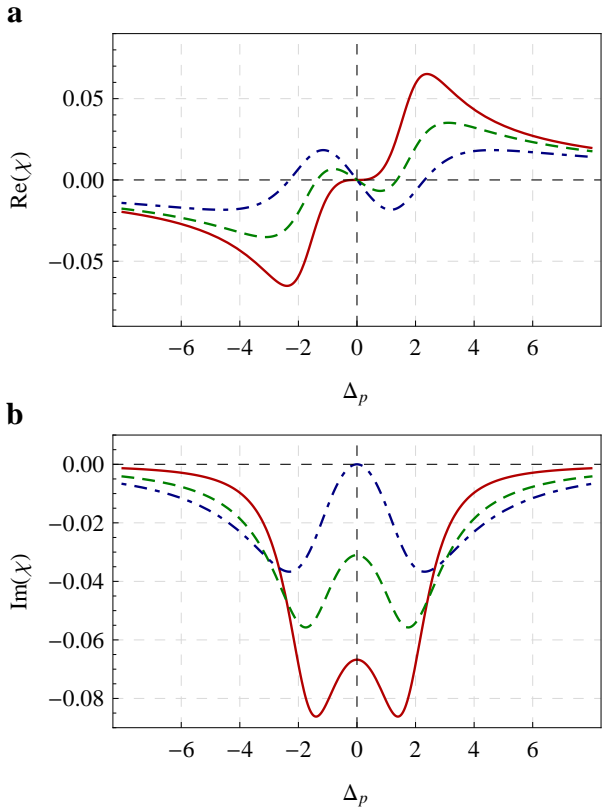}
\caption{Real (a) and imaginary (b) parts of susceptibility versus probe field detuning for the parameters $\gamma = 1$, $r_1=1.5$, $\Omega_c=1.69$, $r_2=0$ (solid), $1.0$ (dashed), and  $2.43$ (dot-dashed).}
\label{fig:F6}
\end{figure}
\begin{figure}[t]
\centering
\includegraphics[width=8cm]{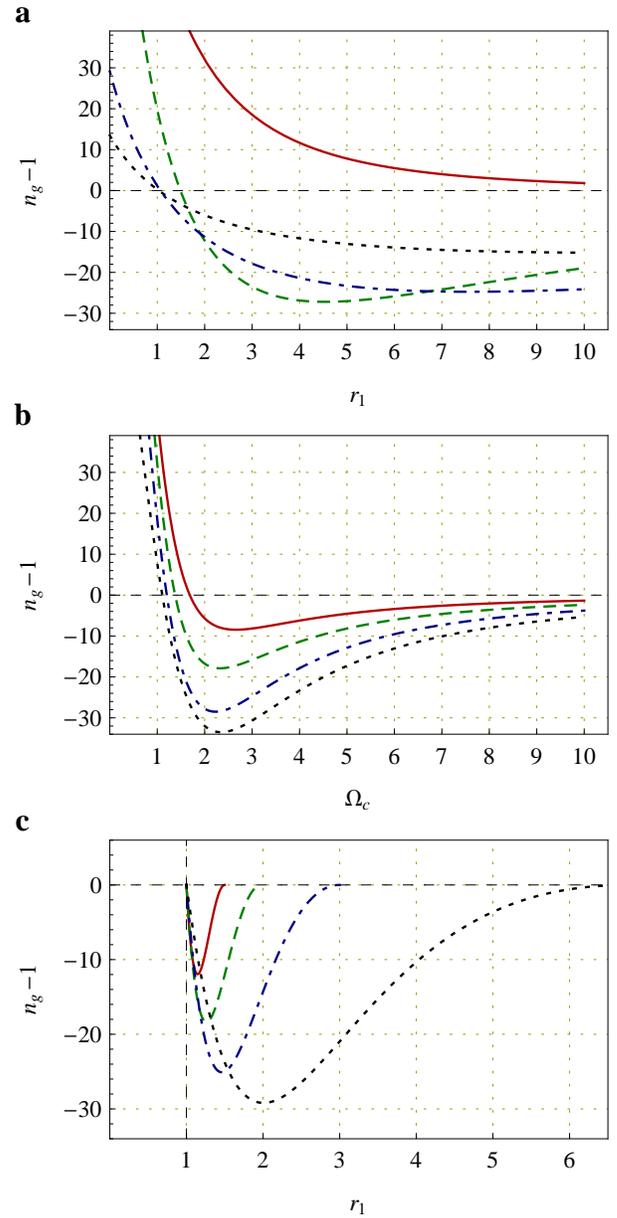}
\caption{(a) The group index, $(n_g-1)$, versus $r_1$ for the parameters $r_2=0.0$, $\Omega_c=1.0$ (solid), $1.7$ (dashed), $4.0$ (dot-dashed), and $6.0$ (dotted). (b) The group index against $\Omega_c$ for the parameters $r_2=0.0$, $r_1=1.5$ (solid), $2.0$ (dashed), $3.0$ (dot-dashed), and $5.0$ (dotted). (c) The group index versus $r_1$ while $r_2 = r_2(r_1 ,\Omega_c)$ defined in Eq. (8) for the parameters $\Omega _c  = 0.5$ (solid), $0.75$ (dashed), $1.0$ (dot-dashed), and $1.7$ (dotted). Other common parameters are $\gamma = 1$ and $\Delta _c  = \Delta _p  = 0$.}
\label{fig:F7}
\end{figure}
Fig.\ref{fig:F4} and Fig.\ref{fig:F5} are presented to clarify how positive
and negative dispersions can occur in the gain region of $\Omega _c$ and $r_1$ plane. By the indirect incoherent pump of
strength $r_1  >\gamma$ (see Eq.\ref{eq:5c}), the inversion of population
between levels $\left| 1 \right\rangle$ and $\left| 3 \right\rangle$ is established. In this situation the system
shows amplification for the probe field. Then there exists a gain deep with positive slope of dispersion
around zero detuning (see the red-solid lines of Fig.\ref{fig:F4}). By applying the coupling field level $\left|1\right\rangle$ is split into two
symmetric dressed levels $\left| {d^+} \right\rangle$ and $ \left| {d^-} \right\rangle$ with frequency
differences of $ \pm \Omega _c$ from the original level. When the coupling field is weak, there still seems to be one gain deep
and the dressed levels are indistinguishable. By strengthening the coupling field, the gain deep is transformed to two gain
deeps (see Fig.\ref{fig:F4}). This is proportional to changing the normal slope of dispersion to anomalous (negative) dispersion which leads to
exceeding the group velocity from $c$  (speed of light in vacuum) or even may cause its becoming negative. When the dressed levels are partly
distinguishable, system may still exhibits a considerable gain around zero detuning, which corresponds to existence of noise in the medium for a
probe pulse with a middle frequency of $\omega _{31}$. By choosing a larger Rabi frequency for the coupling field, the gain becomes lesser and
the anomalous region becomes wider. Then to achieve a transparent medium with anomalous dispersion, i.e. superluminal light propagation, a strong
enough coupling field should be applied while the population is inverted for the probe field
transition \cite{26,27}. If $r_2=0$, as Fig.\ref{fig:F3}a shows, $\Omega_c > 1$ is the sufficient condition to switch the probe
pulse propagation from subluminal to superluminal by application of the indirect incoherent
pump of strength $r_1$. For $\Omega_c\gg1$, $r_1=1$ is the threshold value at which the slope of dispersion changes from positive
to negative. On the other hand, the steepest anomalous slope of dispersion occurs when $ 2 < \Omega _c  < 3$ and a strong enough
indirect incoherent pump, populates level $ \left| 3 \right\rangle$.
Fig.\ref{fig:F5} is presented to see how an increase in $r_1$ causes the gain
deeps to become more distinguishable and consequently the anomalous slope of dispersion become steeper.
In Fig.\ref{fig:F5}a for $r_1=1.2$ (solid line) the slope of dispersion around zero probe field detuning is positive. With increasing the indirect
incoherent pump rate to $r_1=1.5$ (dashed lines), the slope of dispersion becomes zero and it is
negative for $r_1=3.0$ (dash dotted lines). In Fig.\ref{fig:F5}b, the  probe field absorption curves are plotted. It
shows that the subluminal and superluminal light propagation around zero probe detuning is accompanied by a considerable
gain. It is time to investigate the optical response of the medium to populating level $ \left| 2 \right\rangle$ while the
transition $ \left| 1 \right\rangle  \leftrightarrow \left| 3 \right\rangle$ is driven by the coherent field, and the population
between levels $ \left| 1  \right\rangle$ and $ \left| 3 \right\rangle$ is inverted due to the indirect incoherent pump
with a rate $ r_1  > \gamma$.
In Fig.\ref{fig:F6}, the real (a) and imaginary (b) parts of $\chi$ are plotted versus
probe field detuning. The common parameters are $ \Omega _c  = 1.69\gamma$, and $r_1=1.5\gamma$. The first
plots (solid lines), corresponding to $r_2=0$, are presented for comparison only. They show that the slope of
dispersion is almost zero around zero probe field detuning and there is a considerable amount of gain due to the partly
combined gain deeps. By populating level $ \left| 2  \right\rangle$ with rate $r_2=\gamma$, (dashed lines), the slope of
dispersion becomes negative around $ \Delta _p  = 0$. The absorption curve varies according to the probe field detuning, so that the maximum
reduction in gain occurs around zero probe field detuning, while beyond $\Delta _p  =  \pm \Omega _c \gamma$, the gain increases slightly.
By increasing the value of $r_2$ to $2.43\gamma$ (dashed-dotted lines) the region of anomalous dispersion becomes wider and its negative
slope increases in absolute value. The absorption curve undergoes variations similar to those of the previous case ($r_2=\gamma$). The value of the
gain for the latter case ($r_2=2.43\gamma$) vanishes at $\Delta _p = 0$. The proper value of $r_2$ resulting in zero gain around $ \Delta _p  = 0$ can
be obtained by:
\begin{eqnarray}
r_2 (r_1 ,\Omega _c ) = \frac{{r_1^2  + 4r_1 \Omega _c^2  - 2\Omega _c^2  - 1}}{{ - r_1  + 2\Omega _c^2  + 1}}.
\label{eq:8}
\end{eqnarray}
Fig.\ref{fig:F7} displays the effect of incoherent and coherent pump fields on the group index. Fig.\ref{fig:F7}a shows the group index versus
indirect incoherent pump rate $r_1$ for $\Omega_c=1.0$ (solid), $1.7$ (dashed), $4.0$ (dot-dashed), $6.0$ (dotted),
while $r_2=0.0$. Fig.\ref{fig:F7}b displays group index versus coherent pump
field, $\Omega_c$, for $r_1=1.5$ (solid), $2.0$ (dashed), $3.0$ (dash-dotted), and $5.0$ (dotted), while $r_2=0$. In
Fig.\ref{fig:F7}c, the group index is plotted as a function of  $r_1$ for $\Omega_c=0.5$ (solid), $0.7$ (dashed), $1.0$ (dot-dashed), and $1.7$ (dotted)
while the second indirect incoherent pump has rate $ r_2  = r_2 (r_1 ,\Omega _c )$ defined in Eq.(\ref{eq:8}). It shows that
even for $\Omega _c  < 1$ the probe pulse can propagate superluminally while the medium is transparent, if proper
values for $r_1$ and $r_2$ are chosen.

\section{Conclusion}\label{sec:level3}
In conclusion, we have controlled the dispersion and the absorption of a weak probe field in a three-level V-type atomic system. By
application of indirect incoherent pump fields the effect of populating upper levels on optical properties of the atomic medium
in presence of a strong coherent pump field has been investigated and it has been found that linear positive or negative transparent dispersion
could occur between the doublet absorption or gain lines, respectively. Then
the absorption free superluminal light propagation has been established in this system.


\end{document}